\documentclass[letterpaper]{article}

\newcommand{\be}{\begin{equation}}    
\newcommand{\ee}{\end{equation}}

\newcommand{\bean}{\begin{eqnarray}}          
\newcommand{\eean}{\end{eqnarray}}

\newcommand{\bea}{\begin{eqnarray*}}          
\newcommand{\eea}{\end{eqnarray*}}

\newtheorem{Lemma}{Lemma}[section]
\newtheorem{Theorem}{Theorem}[section]
\newtheorem{Corollary}{Corollary}[section]
\newtheorem{Def}{Definition}[section]

\def\F{{\cal F}}
\def\U{{\cal U}}
\def\M{{\cal M}}
\def\O{{\cal O}}
\def\G{{\cal G}}
\def\H{{\cal H}}
\def\Riem{{\cal R}{\it iem}}
\def\c{\chi}
\def\s{\sigma}
\def\vphi{\varphi}
\def\o{\omega}
\def\t{\theta}
\def\b{\flat}
\def\#{\sharp}

\def\l{\lambda}
\def\p{\partial}
\def\d{{\nabla}}
\def\D{{\hat{\nabla}}}
\def\P{{\cal P}_c}
\def\Q{{\cal Q}_c}
\def\C{{\cal C}}
\def\L{{\cal L}_u}
\def\F{{\cal F}_c}
\def\X{{\cal X}_c}
\def\B{{\cal B}}

\def\Star{{\textstyle\star}}

\def\Eq#1{Eq.~(\ref{#1})}

\def\art#1{[\ref{#1}]}

\def\half{{\textstyle {1\over2}}}

\long\def\symbolfootnote[#1]#2{\begingroup%
\def\thefootnote{\fnsymbol{footnote}}\footnote[#1]{#2}\endgroup}

\begin{document}

\begin{center}
\Large{Geometry of Dynamical Systems}
\symbolfootnote[2]
{Published in 
``Proceedings of the Fourth Monash General Relativity Workshop" 
Eds A.Lun, L.Brewin, E.Chow 
(Department of Mathematics, Monash University, 1993), p.38}\\
   \large{A.~A.~Kocharyan}\\
   \small{\sl Department of Mathematics, 
Monash University, Clayton Victoria 3168, Australia\\}
\end{center}

{\bf Abstract.}
We propose a geometrical approach to the investigation 
of Hamiltonian systems on (Pseudo) Riemannian manifolds. 
A new geometrical criterion of instability and chaos is proposed. 
This approach is more generic than well known reduction 
to the geodesic flow. It is applicable to various astrophysical 
and cosmological problems. 

\section{Introduction}

Our main aim is to study global behaviour of dynamical systems
on (Pseudo) Riemannian manifolds using their local,
geometric properties. The guiding principle comes from the well
known example of dynamical system: geodesic flow.

If geodesic flow on Riemannian manifold is given, then to 
work out principal features of the flow,
so-called sectional curvature should be investigated.
And if it turns out that
sectional curvature is negative definite, then one deduces
that geodesic flow is Anosov with hyperbolic geodesics.

As is well known, any Hamiltonian system 
(with the Hamiltonian of the form kinetic plus potential)
is reducible to a geodesic flow.
Therefore, above mentioned procedure may be used for
the new system to get some characteristic features of the original
system. In our opinion it is not 
a proper way to study dynamical systems. 
First of all, it does
not work for all Hamiltonian systems (there must not be turning
points i.e. kinetic energy must be positive), 
then in order to reduce to a geodesic flow
one has to change the metric, Levi-Civita connection,
and time parameter. And since sectional curvature essentially 
depends on the above mentioned values, one must make inverse transforms,
in order to obtain characteristic parameters for the original system,
which is not a trivial problem for typical systems.

Here, in particular, we are looking for a local, geometrical object
(it should be the sectional curvature in the case of the geodesic flow)
for Hamiltonian systems which should characterise instability, 
hyperbolicity of dynamical systems.

We adopt the following approach to the problem, 
which may be considered as an alternative to the approach mentioned
above. Instead of
changing metric, Levi-Civita connection (covariant derivative), 
and time, we change only the derivative. 
We define a new covariant derivative $D_u$ for any flow, which is
determined uniquely and for simple cases, coincides with the
well known derivatives ($\d_u$-covariant derivative, 
$F_u$-Fermi derivative). By means of the derivative
we introduce a coordinate independent equation for
the flow invariant separation vector
(perpendicular to the velocity).
For the geodesic flow it is nothing but the Jacobi equation.
As a by-product we represent any invariant vector field 
as a sum of vectors, which are perpendicular and parallel to the 
velocity respectively.
Then we introduce a geometrical object $\Omega_u$ negativity of which
means hyperbolicity of the corresponding flow. 
This method does work even for systems with turning points
and moreover, which is crucial for the Einstein dynamics,
the geometrical approach applies for Pseudo-Riemannian
manifolds as well.

\section{Geometry of Dynamical Systems}

\subsection{Dynamical Systems}

Let $S$ be a smooth vector field on the tangent bundle $\tau:TM\to M$
of a Riemannian manifold endowed with a Riemannian metric $g$, its
associated Levi-Civita connection $K$ with covariant metric $\d$
(cf. \art{Klingenberg})
$$
  S:TM\to TTM: u\mapsto(u,-\p V)\in Hor(u)\oplus Ver(u)\ ,
$$
where $V:M\to R^1$ and $\p V$ is the gradient vector of V, 
i.e. vector field such that
$$
  \langle dV,X\rangle=g(\p V,X)\ ,\qquad \forall X\in TM\ ,
$$
$Hor(u)$ and $Ver(u)$ are horizontal and vertical
subspaces of $T_uTM$ respectively.
We denote by $u(t)=f^tu_0$ an integral curve of the vector field 
$S$ passing through the initial point $u_0\in TM$ i.e.
$$
  \dot{u}=S(u)\ ,
$$
and $u(0)=u_0$\ .
The group of diffeomorphisms $\{f^t\}$, $t\in R$
$$
  f^t: TM\to TM\ :u_0\mapsto u(t)\ ,
$$
is called a dynamical system or flow of the vector field $S$ on $TM$.
The curve $u(t)=f^tu_0$ is called the flow line or integral curve
starting at $u_0$, $c=\tau\circ u$ is a trajectory.
\begin{Lemma}\label{CurveToTraj}
If $u(t)=f^tu_0$ is an integral curve of the vector field $S$
starting at $u_0$, then $c(t)=\tau\circ u(t)$ is a solution of the
following equation:
\be
   \cases{ \dot{c}=u\ ,\cr
           \d_uu=-\p V\ ,}  \label{Equ}
\ee
determined by $c(0)=\tau\circ u_0$ and $\dot{c}(0)=u_0$.
\end{Lemma}
\begin{Lemma}\label{TrajToCurve}
If $\d_{\dot{c}}\dot{c}=-\p V$, then $\dot{c}(t)=f^t\dot{c}(0)$.
\end{Lemma}

We are interested in the behaviour of the nearby integral curves of the
integral curve $u(t)=f^tu_0$. To deal with that problem,
next we introduce an equation for the invariant vector field along the
curve $c(t)$.
\begin{Lemma}\label{GJacobi}
Invariant vector fields $Z(t)$ along the integral curve $u(t)=f^tu_0$,
i.e.
$$
  Z(t)=Tf^tZ_0\ ,\qquad Z_0\in T_{u_0}TM
$$
are in $1:1$ correspondence with the solutions of the following equation:
\be
  \d^2_uz+\Re_u(z)-H_V(z)=0\ ,\label{Eqz}
\ee
where $\Re_u(z)=Riem(z,u)u$, $H_V(z)=-\d_z\p V$.
The correspondence is given by
$$
  z(t)\leftrightarrow Z(t)=(z(t),\d_uz(t))\in Hor(u(t))\oplus Ver(u(t))\ .
$$
\end{Lemma}
Note that one might consider \Eq{Eqz} as
an equation obtained from \Eq{Equ} merely by covariant 
differentiation with respect to $z$.

Thus we arrive at the following system of variational equations
\be
   \cases{ \dot{c}=u\ ,\cr
           \d_uu=-\p V\ ,\cr
           \d^2_uz+\Re_u(z)-H_V(z)=0\ .}   \label{dV}
\ee
It is worth mentioning that the vector $z$ describes the
motion of points (but not integral curves) of the nearby 
integral curves of $u(t)$.
However, in most cases it is important to know the behaviour of nearby
integral curves (which means $z$ is orthogonal to velocity $u$)
with the same integrals.

If $V=0$ then the flow corresponding to $S$ is called
a geodesic flow and \Eq{dV} reads as
$$
   \cases{ \dot{c}=u\ ,\cr
           \d_uu=0\ ,\cr
           \d^2_uz+\Re_u(z)=0\ ,} 
$$
the last equation is the Jacobi equation.

We say that the Jacobi field $z$ (solution of the Jacobi equation)
is orthogonal to $u$ if the following conditions hold
$g(u,z)=0$ and $z(E)=g(u,\d_uz)=0$ along integral curve.
One can represent any Jacobi field as follows:
\be
  z=n+[n(E)t+g(u,z)|_0]u\ , \label{JacobiField}
\ee
where $n$ is orthogonal to $u$, i.e. $g(u,n)=g(u,\d_un)=0$\ .
One may prove (see \art{KobNom}) that in the case of the geodesic 
flow $n$ is orthogonal to $u$ for all $t$ if it is so at $t=0$.
So if $z(0)=n(0)$ and $n(E)=0$ then $z(t)=n(t)$ for arbitrary $t$.
It is not true for a general flow.

It is an important property of the geodesic flow, because it means
that the equation for $n$ is the same as for $z$ and one should
only carefully choose the initial conditions.

As is well known (see \art{Arnold}) by changing the 
time parameter $t$, metric $g$, and linear connection $K$ 
(covariant derivative $\d$) one can reduce a Hamiltonian system 
to a geodesic one.
\begin{Theorem} 
On $TM|_E=\{u\in TM\ |\ g(u,u)=2(E-V(\tau\circ u))\ne0\}$\ ,
there exist
\begin{description}
   \item[G1.] $s=\varphi(t)$ parameter such that $\dot\varphi=\sqrt2|E-V|$\ ,
   \item[G2.] Metric $\hat{g}=|E-V|g$\ ,
   \item[G3.] Linear Levi-Civita connection $\hat{K}$ with a
   covariant derivative
$$
  \D_XY=\d_XY+X(\phi)Y+Y(\phi)X-g(X,Y)\p\phi\ ,
$$
\end{description}
where $\phi=\half\ln|E-V|$\ ,
such that instead of \Eq{dV} one has
$$
   \cases{ \gamma'=v\ ,\cr
           \D_vv=0\ ,\cr
           \D^2_vn+\hat{\Re}_v(n)=0\ ,} 
$$
where $\gamma'=\frac{\textstyle d\gamma}{\textstyle ds}$,
$\hat{R}iem$ is the Riemannian tensor of the linear connection
$\hat{K}$, 
$$
  \half g(u,u)=E-V\quad\Longleftrightarrow\quad\hat{g}(v,v)=sgn(E-V)\ .
$$
\end{Theorem}
So any Hamiltonian system can be reduced to a geodesic one.
But this procedure has several demerits:

1) time parameter $t$ changes to affine parameter $s$, to
   recover $t$ one has to find function $\varphi$;

2) metric $g$ changes to $\hat{g}$ it changes Levi-Civita
   linear connection and the measure of length of vectors.

But as is well known local stability/instability
does depend on metric and time. And in order to reduce a
Hamiltonian system to a geodesic one we have changed the both concepts.
Therefore, to make a conclusion about stability/instability of original
system one has to make the inverse transform. Which turns out to
be a complicated one. To overcome the problem we suggest an alternative
method. For this purpose we will introduce a new covariant derivative
for dynamical systems and derive desired equation by help of the
derivative.

Now we will see how to overcome these difficulties.

\subsection{Covariant Derivative of Dynamical System}

Let us introduce a new covariant derivative along curve $c$.
Given a smooth curve $c: R\to M$, $c\in C^{\infty}(R,M)$,
with the velocity vector $u=\dot{c}=Tc.\frac{\textstyle d}{\textstyle dt}$\ .
We denote the set of all smooth vector fields (functions) along curve $c$ 
by $\X$ ($\F$) respectively.
First we define the set of curves $\C$ along which one may define
the covariant derivative.
\begin{Def}
We say that $c\in\C$ if there exists unique smooth $(1,1)$ type 
tensor field $\Q$ along $c$ such that
$$
  \Q=\frac{u\otimes u^\b}{g(u,u)} \qquad
  \mbox{ for all } u \mbox{ such that } g(u,u)\ne0\ .
$$
\end{Def}
Clearly, if $g(u,u)\ne0$ along a curve $c$ then $c\in\C$ .
Another important subset of $\C$ is established by the following Lemma.
\begin{Lemma}
If $c$ is a non-constant solution of the \Eq{Equ} then $c\in\C$ and
$$
 \Q=\cases{ \frac{\textstyle u\otimes u^\b}{\textstyle \|u\|^2}   
                 & if  $g(u,u)\ne0$, \cr\cr
            \frac{\textstyle \p V\otimes dV}{\textstyle \|dV\|^2} 
                 & if  $g(u,u)=0$.}
$$
\end{Lemma}
Let $\P=\delta-\Q$\ .
\begin{Def}
Given $c\in\C$ and $X\in\X$.
We will say that $X\in\X^D$ i.e. $X$ is a $D$-differentiable along $c$
if there exists unique smooth vector field $D_uX\in\X$ such that
if $g(u,u)\ne0$ then
\be
  D_uX=\P\d_u\P X+\Q\L\Q X\ .\label{Du}
\ee
\end{Def}
Obviously, if $g(u,u)\ne0$ along $c$ then $\X^D=\X$\ .
If $g(X,u)=fg(u,u)$, where $f\in\F$ then $X\in\X^D$\ . 
In particular, $u\in\X^D$. 
Besides, if $g(X,u)=0$ then $X\in\X^D$\ .
$D_u:\X^D\to\X\ :X\mapsto D_uX$ is a derivative (see \art{KobNom}).

Some easy-to-check properties of $D_u$:
\begin{description}
 \item[1.] $D_uX=\Q\L X$  if $\Q X=X$\ $(\P X=0)$.
 \item[2.] $D_uX=\P\d_uX$ if $\P X=X$\ $(\Q X=0)$.
 \item[3.] $D_u=\d_u$ if $\d_uu=0$ i.e. $c$ is a geodesic.
 \item[4.] $D_u=F_u$ if $g(u,u)=const\ne0$. $F_u$ is the Fermi derivative.
 \item[5.] $D_uX=\P\d_u\P X+\L\Q X$\ .
\end{description}

The derivative along $c$ can be naturally extended from vector fields to
arbitrary tensor fields as follows:

\begin{description}
   \item[D1.] $D_u$ is a linear mapping of tensor fields of any type 
      along $c$ to tensor fields of the same type, 
      which commutes with contractions;
   \item[D2.] $D_u(T_1\otimes T_2)=D_uT_1\otimes T_2+T_1\otimes D_uT_2$;
   \item[D3.] $D_uf=u(f)={\frac{\textstyle df}{\textstyle dt}}$,
              for any function $f$.
\end{description}
\begin{Theorem}
There exists exactly one derivative determined
by (\ref{Du}) with 1-2 and D1-D3 properties.
\end{Theorem}
\begin{Def}
Given $c\in\C$ and $X\in\X$. 
We say that $X$ is a $D$-parallel vector field along $c$ if $D_uX=0$.
\end{Def}
\begin{Theorem}
For any $Y\in T_{c(0)}M$ (if $g(u,u)_{|0}=0$ then $\Q Y=0$) 
there exists unique parallel vector field 
$X$ along $c$ such that $D_uX=0$ and $X(0)=Y$.
\end{Theorem}
If $\Q X_{|0}=0=\Q Y_{|0}$ and $D_uX=0=D_uY$ then
$g(X,u)=0$, $g(Y,u)=0$, and $g(X,Y)=const$ along curve.
Notice that $D_uu=0$.

Now we derive the desired equation for $n=\P z$ by terms of the 
covariant derivative.
\begin{Theorem}
Given $c$ which satisfies the \Eq{Equ}. Then any solution of the \Eq{Eqz}
has the following form (we consider limit-values of all objects at $g(u,u)=0$)
\be
  z=n-\left[\int\frac{-2n(V)+n(E)}{2(E-V)}dt\right]u\ ,
  \label{z}
\ee
where $n=\P z$ satisfies the following equation
\bea
&&  D_u^2n+\Omega_u(n)=\frac{n(E)}{E-V}\P\p V\ ,\\
&&  \Q n|_{t_0}=\Q D_un |_{t_0}=0\ ,
\eea
where
\bea
&& \Omega_u(n)=\Re_u(n)+\P\left[-H_V(n)+\frac{3n(V)}{2(E-V)}\p V\right]\ ,\\
&& E=\half g(u,u)+V=const \qquad\mbox{along integral curve,}\\
&& n(E)=z(E)=g(u,\d_uz)+z(V)=const \qquad\mbox{along integral curve.}
\eea
\end{Theorem}

Let us mention that if $V=0$, and $E=\half$ along the integral curves
(geodesics with $g(u,u)=1$) then decomposition of $z$ \Eq{z} coincides
with the Jacobi field decomposition \Eq{JacobiField}.
If we consider all integral curves of fixed energy then the following
claim holds.
\begin{Corollary}
If $\Q z |_{t_0}=0$, $z(E)=0$,
then
$$
  z=n-\left[\int_{t_0}^t\frac{n(V)}{E-V}d\tau\right]u\ , 
$$
where $n=\P z$ satisfies the following equation
\bean
&&  D_u^2n+\Omega_u(n)=0\ ,\label{Eqn}\\   
&&  \Q n |_{t_0}=\Q D_un |_{t_0}=0 \ .\nonumber
\eean
\end{Corollary}
\begin{Corollary}
If $g(u,u)\ne0$ (there is no turning point) and $n(E)=0$ then 
$$
\begin{array}{lr}
   \cases{ D_u^2n+\Omega_u(n)=0\ ,\cr
           g(u,n)|_0=g(u,D_un)|_0=0\ ,}
&\Longleftrightarrow\quad
   \cases{ \D_v^2n+\hat{\Re}_v(n)=0\ ,\cr
           \hat{g}(u,n)|_0=\hat{g}(u,\D_vn)|_0=0\ .}
\end{array}
$$
\end{Corollary}
One can derive an equation for the length of $n$ from \Eq{Eqn}.
Let $n=\ell\xi$, where $\|n\|=\ell$ and therefore 
$g(\xi,\xi)=1$ and $g(\xi,u)=0$. Then
\be
  \ddot{\ell}+[\omega_u(\xi,\xi)-g(D_u\xi,D_u\xi)]\ell=0\ ,\label{l}
\ee
where
\bea
&& \omega_u(\xi,\xi)=g(\Omega_u(\xi),\xi)
              =K(\xi,u)-h_V(\xi,\xi)+\frac{3[\xi(V)]^2}{2(E-V)}\ ,\\
&& K(\xi,u)=g(\Re_u(\xi),\xi)=Riem(u,\xi,u,\xi)\ ,\\
&& h_V(\xi,\xi)=g(H_V(\xi),\xi)\ .
\eea
And
$$
 D^2_u\xi+2\l D_u\xi+(\dot{\l}+\l^2)\xi+\Omega_u(\xi)=0\ ,\label{Dxi}
$$
where $\l=\dot{\ell}/\ell$.

It is worth noting that $\Omega_u(\xi)$ and $\omega_u(\xi,\xi)$
are smooth functions with respect to their all arguments ($x,u,\xi$) 
along integral curves, even at turning points. And if $u(0)=0$ then
$$
  \omega_0(\xi,\xi)=\lim_{t\to0}\omega_u(\xi,\xi)=\|dV\|k(\xi,\xi)\ ,
$$
where $k$ is the second fundamental form of
the $V=E$ submanifold in $M$ and $\langle dV,\xi\rangle =0$.
\begin{Theorem}
If $SM_E=\{u\in TM|\half g(u,u)+V(\tau\circ u)=E\}$ is a compact manifold,
$g(u,u)+\|dV\|^2\ne0$ (i.e. there is no singular point), and 
\be
  \omega_u(\xi,\xi)<0\ , \label{Negative}
\ee
where $u\in SM_E$, $g(\xi,\xi)=1$, and $g(u,\xi)=0$
then the dynamical system (\ref{Equ}) is Anosov.
\end{Theorem}
Condition (\ref{Negative}) may be fulfilled unless dynamical
system has an integral on $SM_E$.
One may average $\omega_u(\xi,\xi)$ by replacing
$\xi\otimes\xi$ with 
$$
  \xi\otimes\xi=\frac{1}{d-1}\left(g-\frac{u\otimes u}{2(E-V)}\right)\ ,
$$
to get
$$
  \B(u)=\frac{1}{d-1}\left[Ric(u,u)-\Delta V+\frac{h_V(u,u)}{g(u,u)}
               +\frac{3|u\wedge\p V|^2}{g(u,u)^2}\right]\ ,
$$
where
$$
  |X\wedge Y|^2=g(X,X)g(Y,Y)-g(X,Y)^2\ ,
$$
and
$$
  \B_0(x)=\lim_{t\to0}\B(u)
        =\frac{1}{d-1}\left[-\Delta V+\frac{h_V(\p V,\p V)}{\|dV\|^2}\right]\ .
$$
One may speculate that a solution of the \Eq{l} in average is greater than
a solution of the following ``evolution equation" (cf. \art{Arnold}, \art{Book})
\be
  \ddot{\ell}(t)+\B(t)\ell(t)=0\ ,\label{meanl}
\ee
where $\B(t)=\B(u(t))$. Notice that \Eq{meanl} is an exact if $d=2$.

\subsection{Pseudo-Riemannian Manifolds}

The impossibility of direct application of the results 
developed for Riemannian manifolds 
is evidently connected with 
the Pseudo-Riemannian signature of the metric,
e.g. it is easy to see that two close geodesics can diverge 
in Pseudo-Riemannian manifold, while the length of the
separation vector may remain close to zero or even negative. Therefore not 
only new criteria are needed here but one should also redefine the 
stability properties themselves.

Consider a dynamical system on Pseudo-Riemannian manifold $M$
(see \art{Hyp}).
Let $c$ be a trajectory passing through a point $c(0)=m\in M$
and $\{E_a^0\}$ $a=1,\ldots,d$ basis for $T_mM$. 
One can propagate $\{E_a^0\}$ along $c$ so that
$$
  D_uE_a=0\ ,\quad E_a(0)=E_a^0\ , \qquad \forall a=1,\ldots,d\ ,
$$
and get a basis $\{E_a\}$ along the curve $c$ for any $T_{c(t)}M$. 
We will call it $D$-basis.
Any vector $X\in T_{c(t)}M$ can be presented by means of the 
$D$-basis $\{E_a\}$
$$
  X(t)=X^a(t)E_a\ .
$$
The expression
$$
  g_E(X,Y)=\sum_aX^aY^a
$$
defines $E$-metric on $TM$. 
Length of the vector $X$ with respect to the basis $\{E_a\}$ is defined
to be
$$
  \|X\|_E=(g_E(X,X))^{1/2}=\left(\sum_a(X^a)^2\right)^{1/2}.
$$
Let $\{E_{a'}\}$ be another $D$-basis along the same curve $c$.
Then a non-singular matrix $\Phi$ exists, such that
$$
  E_a^0=\sum_{b'}\Phi_a{}^{b'}E_{b'}^0\ .
$$
So far as both $\{E_a\}$ and $\{E_{a'}\}$ being $D$-bases are
$D$-parallel transported along $c$, this relation must 
hold for constant $\Phi$. Therefore
$$
  E_a=\sum_{b'}\Phi_a{}^{b'}E_{b'}\ .
$$
Thus we have 
$$
  X^{b'}(t)=\sum_{a}\Phi_a{}^{b'}X^a(t)\ .
$$
Since $\Phi$ is non-singular then
$$
  \sum_{b'}(X^{b'})^2>0\ ,
$$
for if $\sum\limits_{b'}(X^{b'})^2=0$ we get
$$
  \sum_{a}\Phi_a{}^{b'}X^a(t)=0\ ,
$$
it is a contradiction with non-singularity of $\Phi$.
Therefore we have a positive defined matrix 
$$
  \Psi_{ac}X^aX^c=\sum_{b'}\Phi_a{}^{b'}\Phi_c{}^{b'}X^aX^c
                 =\sum_{b'}(X^{b'})^2>0\ .
$$
Space of all $X$ with $\|X\|_E^2=1$ is a compact space, therefore
there are positive constants $\alpha$ and $\beta$ 
(depending only on basis) such that
$$
 0<\alpha\sum_{a}(X^a)^2\leq\sum_{b'}(X^{b'})^2\leq\beta\sum_a(X^a)^2
$$
or
$$
 0<\alpha\|X\|_E^2\leq\|X\|_{E'}^2\leq\beta\|X\|_E^2\ .
$$
For the vector field $n$: $n=n^aE_a$, $D_un=\dot{n}^aE_a$, and
$$
  \ell_E(n)=(\|n\|_E^2+\|D_un\|_E^2)^{1/2}\ .
$$
Now we can define stability, hyperbolicity, etc. of integral curves.
\begin{Def} 
Integral curve $u$ is called to be linear stable if
$\forall\varepsilon>0$, $\exists\delta(\varepsilon)>0$ such that
$\ell_E(n(0))<\delta(\varepsilon)\ \Rightarrow \ell_E(n(t))<\varepsilon\ ,$
for all $t\geq0$, where $n$ is a solution of the \Eq{Eqn}.

If $u$ is not linear stable it is called linear unstable.
\end{Def}
One can readily verify that the definition does not depend on the choice
of $D$-basis $\{E_a\}$.
Lyapunov characteristic exponents $\chi$ for an integral curve $u$ is
defined as follows
$$
  \chi(u,n)
  =\limsup_{t\to\infty}\frac{\ln\ell_E(n)}{t}\ .
$$
Evidently $\chi$ does not depend on choice of $\{E_a\}$ either.
One can give a definition of hyperbolicity as well (see \art{Hyp}).

Let $n=\ell\xi$, where $\|n\|_E=\ell$ and therefore $\|\xi\|_E=1$, then
(cf. \Eq{l})
$$
  \ddot{\ell}+[g_E(\Omega_u(\xi),\xi)-g_E(D_u\xi,D_u\xi)]\ell=0\ ,
$$
where
$$
 g_E(\Omega_u(\xi),\xi)=g_E(\Re_u(\xi),\xi)-g_E(H_V(\xi),\xi)
                       +\frac{3g(\xi,\p V)g_E(\xi,\p V)}{2(E-V)}\ .
$$

\section{Einstein Dynamics on Superspace}

\subsection{World, Universe, Superspace}

The set of $d$-dimensional Universes will be described as follows
(see \art{Super}). 
We assume, that the Universe is closed (compact and without boundary).

By $\M^{d+1}$ we denote the set of all $d+1$-dimensional, smooth 
(from the class $C^r$, $r>2$), oriented, compact manifolds ($d>1$),
\bea
& \M^{d+1}=\{M^{d+1}\}=\{&\mbox{all } d+1 \mbox{-dimensional smooth,}\\
&                        &\mbox{oriented, compact manifolds \ }\}\ .
\eea
World (i.e. spacetime with material fields) will mean the following triad:
$$
    (M^{d+1},g(M),\Phi(M))\ ,
$$
where $M^{d+1}\in\M^{d+1}$, and $g(M)$ is a smooth Riemannian metric on
$M^{d+1}$, $\Phi(M)$ is a smooth scalar field. Denote the set of worlds by 
$W^{d+1}$,
$$
  W^{d+1}=\{w\}=\{(M^{d+1},g(M),\Phi(M))\}=\{(M_w,g_w,\Phi_w)\}\ .
$$
Let $\F(M^{d+1})$ be a set of smooth functions on $M^{d+1}$ without singular
critical points (Morse's function) such that $f\in\F(M^{d+1})$ if
\bea
&& f:M^{d+1}\to S^1=[0,2\pi]/\{0,2\pi\}\ ,\\
&& f\in C^r(M^{d+1})\ ,\\
&& \mbox{ if } \p M^{d+1}\ne\emptyset 
   \mbox{ then }f(\p M^{d+1})=\p[f(M^{d+1})]\ ,
\eea
where $\p N$ is the boundary of the manifold $N$\ .
For every $c\in S^1$, $f\in\F(M^{d+1})$ we denote
\bea
&& f_c[M^{d+1}]=\{x|x\in M^{d+1}, f(x)=c\}\ ,\\
&& Y\{f_c[M^{d+1}]\}=\{x|x\in f_c[M^{d+1}], df|_{f_c[M^{d+1}]}(x)=0\}\ .
\eea
For given $w\in W^{d+1}$, $f\in\F(M_w)$, and $c\in S^1$ we have the 
following triad:
$$
  u(w,f,c)=(f_c[M_w],g,\phi)\ ,
$$
where $g$ is the metric induced on $f_c[M_w]$ by $g_w$, 
$\phi=\Phi_w|_{f_c[M_w]}$. Universes (i.e. space with material fields) are 
members of the set $\U^d$,
\bea
&& \U^d=\bigcup_{w\in W^{d+1}}\bigcup_{f\in\F(M_w)}
        \bigcup_{c\in S^1}u(w,f,c)\ ,\\
&& \U^d=\{u\}=\{(T_u,g_u,\phi_u)\}\ .
\eea
According to Morse's theory 
$$
  \U^d=\Sigma^d\cup\Omega^d\cup\{\emptyset\}\ ,
$$
where $\Sigma^d$ is the set of all $d$-dimensional smooth, oriented,
closed manifolds with a smooth Riemannian metric and a smooth scalar 
field on them. We will consider the empty set $\emptyset$ as a
trivial manifold.

At Morse's reconstructions $\Omega^d$ are critical sets with a given metric
and scalar fields.

In order to construct the set $\U^d$ as a topological space one needs
to introduce a topology, a system of open sets. We will do it as follows:
first we define the set of ``smooth" curves on $\U^d$.

A mapping $\l:(0,1)\to\U^d$ is called a ``smooth" one in $\U^d$, if
$\exists w\in W^{d+1}$ and $\exists f\in\F(M_w)$ such that
$$
  \l_{(w,f)}(\tau)=u(w,f,\tau),\qquad \tau\in(0,1)\subset S^1\ .
$$
$\O$ is the strongest topology on the set $\U^d$, such that every 
``smooth" curves from the following set
$$
  \bigcup_{w\in W^{d+1}}\bigcup_{f\in\F(M_w)}\l_{(w,f)}(\cdot)
$$
is continuous on $(0,1)$.

By Superspace we mean an $\U^d$ endowed with the topology $\O$
$(\U^d,\O)$. We denote that space by $\U^d$ again.

The space of Universes with given manifold $M$ we denote by $\U^d_M$,
$$
 \U^d_M=\{u\in\U^d| T_u=M\}\ .
$$

\subsection{Geometry of Superspace}

Let us fix any $M\in\Sigma^d$ and consider $\U^d_M$. In this case
$\U^d_M$ is the space of all smooth Riemannian metrics and smooth
scalar fields on $M$. It is known that there exists smooth Banach
structure on such spaces. 

We will introduce a metric on $\U^d_M$ such that kinematical part 
of the Hamiltonian given by ADM formalism could be expressed by that
metric. So we have the following metric on $\U^d_M$
$$
  \G[g,\phi](k,\c;h,\t)=\int_M[-tr(k)tr(h)+tr(k\times h)+\c\t]
                              N^{-1}d\mu(g)\ ,
$$
where $N$ is a positive function $N: M\to R_+=\{x\in R | x>0\}$ and
\bea
&& (g,\phi)\in\U^d_M\ ,\qquad (k,\c), (h,\t)\in T_{(g,\phi)}\U^d_M\ ,\\
&& d\mu(g)=(\det g)^{1/2}dx^1\wedge\ldots\wedge dx^d\ ,\\
&& tr(k)=g^{ab}k_{ab}\ ,\qquad (k\times h)_{ab}=k_{ac}g^{cd}h_{db}\ .
\eea
The metric has an inverse metric $\G^{-1}$ and
$$
  \G^{-1}[g,\phi](\pi,p;\varrho,q)=
  \int_M\left(-\frac{1}{d-1}tr(\pi')tr(\varrho')
                               +tr(\pi'\times\varrho')+p'q'\right)Nd\mu(g)\ ,
$$
where
$$
 (\pi,p), (\varrho,q)\in T_{(g,\phi)}^*\U^d_M\ ,\qquad 
 \pi=\pi'd\mu(g),\ldots,q=q'd\mu(g)\ .
$$
By means of this metric one can map $T\U^d_M$ on $T^*\U^d_M$ and 
vice versa--$T^*\U^d_M$ on $T\U^d_M$. These mappings are of the 
following form:
\bea
&& \G^{\b}:T\U^d_M\to T^*\U^d_M\ ,\\
&& \G^{\b}[g,\phi](k,\c)=(-tr(k)g^{-1}+k^{-1},\c)d\mu(g)
                        =(-tr(k)g^{ab}+k^{ab},\c)d\mu(g)\ ,\\
&& \G^{\#}:T^*\U^d_M\to T\U^d_M\ ,\\
&& \G^{\#}[g,\phi](\pi,p)=\left(-\frac{1}{d-1}tr(\pi')g+\pi'^{\b},p'\right)
                         =\left(-\frac{1}{d-1}tr(\pi')g_{ab}+\pi'_{ab},p'\right).
\eea
As on a manifold of finite dimension, on $\U^d$ there exists the unique
Levi-Civita connection, i.e. a Riemannian one without torsion.
In this case as it is shown in \art{Super}, 
\bea
\lefteqn{\Gamma[g,\phi](k,\c;\o,\vphi;h,\t)
         =\frac{1}{4}\int_M\{-tr(k)tr(\o)tr(h)}\\
&& +3tr(k)tr(\o\times h)+tr(\o)tr(k\times h)+tr(h)tr(k\times\o) \\
&& -4tr(k\times\o\times h)+tr(h)\c\vphi+tr(\o)\c\t-tr(k)\vphi\t\}
      N^{-1}d\mu(g)\ ,
\eea
and
\bea
\lefteqn{\Riem[g,\phi](\o,\vphi;k,\c;l,\s;h,\t)}\\
&&=\int_M\{(1/4)tr[(\bar{k}\times\bar{\o}-\bar{\o}\times\bar{k})
            \times(\bar{h}\times\bar{l}-\bar{l}\times\bar{h})]\\
&&+\kappa^2[tr(\bar{\o}\times\bar{l})tr(\bar{k}\times\bar{h})
           -tr(\bar{h}\times\bar{\o})tr(\bar{k}\times\bar{l})\\
&&+tr(\bar{\o}\times\bar{l})\c\t-tr(\bar{\o}\times\bar{h})\c\s
+tr(\bar{k}\times\bar{h})\vphi\s-tr(\bar{k}\times\bar{l})\vphi\t]\}N^{-1}d\mu(g)\ ,
\eea
where
$$
 \kappa^2=\frac{d}{16(d-1)},\qquad \bar{\o}=\o-\frac{tr(\o)}{d}g\ .
$$

Consider now the dynamics with the following Hamiltonian in $M$-superspace
$$
 \H[g,\phi,\pi,p]=\half\G^{-1}[g,\phi](\pi,p;\pi,p)+V[g,\phi]\ ,
$$
where
\bea
&& V[g,\phi]=\int_M\{-R(g)+\half\|d\phi\|^2+F(\phi)\}d\mu(g)\ ,\\
&& \|d\phi\|^2=g^{ab}\phi_{|a}\phi_{|b}\ .
\eea
The dynamics given by this Hamiltonian 
$$
  \d_uu=-\p V\ ,
$$
where $\d$ is a covariant derivative on the $M$-Superspace,
corresponds to Einstein's equations together with the 
following constraints equations,
\bea
&& \H[g,\phi;\pi,p]=0\ ,\\
&& {\cal P}_b[g,\phi;\pi,p]=-2\pi^{a}{}_{b|a}+p\phi_{|b}=0\ .
\eea
These equations constrain only initial conditions, but not dynamics.

Now to investigate stability/instability of the solutions of the 
Einstein's equations the following objects should be used \art{WiP},
(in order to derive $\Omega_u$)
$$
  \p V=N(\{S^{\Star}-c\cdot Fg\}N+Hess(N),
              \left\{F'+\Delta\phi\right\}-g(\p\phi,\p N))\ 
$$
and
$$
 -H_V(h,\c)=\d_{(h,\c)}\p V={\cal P}(h,\c)+{\cal D}(h,\c)+{\cal N}(h,\c)\ ,
$$
where
\bea
{\cal P}(h,\c)=N^2\left(
  \half r(h)+c[\{tr(S\times h)-F'\c\}g-(tr(S)+F)h],\right.\\
  -tr(Hess(\phi)\times h)+F''\c)\ ;\\
{\cal D}(h,\c)=N^2\left(
  \left[\half(\Delta h-\alpha\delta h+Hess(tr(h))
                             -d\phi\otimes dtr(h)\right]^\Star\right.,\\
  \left.-g(\p\phi,\delta h+\half\p tr(h))+\Delta\c\right)\ ;\\
{\cal N}(h,\c)=N(-D_g(Hess(N)).h,-g(\p\c,\p N))\ ,
\eea
\bea
& c=\big(2(d-1)\big)^{-1}\ ,\quad & \p f=g^{ab}f_{|b}\ ,\\
& T^\Star=T-c\cdot tr(T)g\ ,      & S=Ric-\half d\phi\otimes d\phi\ ,\\
& Hess(f)=-f_{|a|b}\ ,            & \Delta f=tr(Hess(f))=-g^{ab}f_{|a|b}\ ,\\
& \delta h=-h^{ab}{}{}_{|b}\ ,    & \alpha(X)={\cal L}_Xg=X_{a|b}+X_{b|a}\ ,
\eea
$$
r(h)_{ab}=-Riem_a{}^c{}_b{}^dh_{cd}-Riem_b{}^c{}_a{}^dh_{cd}
          +Ric_a{}^ch_{cb}+Ric_b{}^ch_{ca}\ .
$$
In further investigation \art{WiP} we will employ suggested
method to the problem of stability of cosmological solutions.

\section{Conclusions}

Thus, we have proposed a geometrical approach to the investigation of
Hamiltonian systems on (Pseudo) Riemannian manifolds. We have derived
the equation for the normal to the velocity of the flow invariant 
vector \Eq{Eqn} in terms of the new covariant derivative.
We have introduced tensor $\Omega_u$ which characterises stability, 
instability, hyperbolicity of the integral curves of Hamiltonian systems.
This approach is valid for any Hamiltonian system, while well known
reduction to the geodesic flow requires some more conditions.
Besides, there is no need to change metric and time parameter.
This allows obtaining characteristic parameters by
means of physical values.
As an important application we will consider in details the behaviour
of the N-body gravitating systems and cosmological solutions of the 
Einstein's equations elsewhere \art{WiP}.


\begin{thebibliography}{00}


\bibitem{}
\label{Klingenberg}
Klingenberg,~W. Lectures on Closed Geodesics, Springer-Verlag, 1978.

\bibitem{}
\label{KobNom}
Koboyashi,~S., Nomizu,~K. Foundation of Differential Geometry, 
                         {\bf vols 1,2}, Interscience-Publ., 1963, 1969.

\bibitem{}
\label{Arnold}
Arnold,~V.I. Mathematical Methods of Classical Mechanics, Springer-Verlag, 
             1978.

\bibitem{}
\label{Book}
Gurzadyan,~V.G., Kocharyan,~A.A. Paradigms of the Large-Scale Universe,
				Gordon and Breach, 1994.


\bibitem{}
\label{Hyp}
Gurzadyan,~V.G., Kocharyan,~A.A. Hyperbolicity in Pseudo-Riemannian spaces,
                                 YerPhI-920(71), 1986.

\bibitem{}
\label{Super}
Kocharyan,~A.A. Instability in Superspace 
                {\it Comm.~Math.~Phys.} {\bf 143}, pp.~27-42, 1991.

\bibitem{}
\label{WiP}
Kocharyan,~A.A. (work in progress).

\end{thebibliography}
\end{document}